\documentclass[11pt,twoside]{article}

\usepackage{asp2006}
\usepackage{epsfig} 
\usepackage{psfig}
\usepackage{lscape}

\usepackage{graphicx} 

\begin{document}
\title{Exoplanet research with \textit{SAFARI}:\\
A far-IR imaging spectrometer for SPICA}   
\author{Javier R. Goicoechea$^1$, Kate Isaak$^2$ \& Bruce Swinyard$^3$\\
on behalf of the SAFARI consortium.}   
\affil{$^1$Centro de Astrobiolog\'{\i}a. CSIC/INTA. Madrid, Spain.\\
$^2$Cardiff University, UK.\\$^3$Rutherford Appleton Laboratory, Chilton, Didcot, UK.}    

\begin{abstract} 
The far-IR spectral window plays host to a wide range of both spectroscopic 
and photometric diagnostics with which to open the protoplanetary disks and exoplanet 
research to wavelengths completely blocked by the Earth’s atmosphere.
These include some of the key atomic (\textit{e.g.,}~oxygen) and molecular 
(\textit{e.g.,}~water) cooling lines, the dust thermal emission, the water 
ice features as well as many other key chemical tracers. Most of these features 
can  not be observed from ground-based telescopes but play a critical diagnostic in 
a number  of key areas including the early stages of planet formation and potentially 
exoplanets. The proposed Japanese-led IR space telescope SPICA, 
with its 3.5m cooled  mirror will be the next step in sensitivity after \textit{Herschel}. 
SPICA has been recently  selected to go to the next stage of the 
\textit{ESA's Cosmic Vision~2015-2025} process.
This contribution summarizes the design concept behind \textit{SAFARI}: 
an imaging far-IR spectrometer covering the $\sim$30-210\,$\mu$m waveband that is one of 
a suite of instruments for SPICA; it also highlights some of the science questions 
that it will be possible to address in the field of exoplanets and planet formation.

\end{abstract}

\section{Introduction}  

The study of exo-planets (EPs) requires many different approaches across the full wavelength 
spectrum to both discover and characterize the newly discovered objects in order that
 we might fully understand the prevalence, formation and evolution of planetary systems.  
The mid infrared (MIR) and far infrared (FIR) spectral regions
are especially important in the study of planetary atmospheres as it spans both the peak of 
thermal emission from the  majority of EPs thus far discovered (up to $\sim$1000 K) 
and is particularly rich in molecular  features that can uniquely identify the chemical 
composition,  from protoplanetary disks to planetary atmospheres and trace the
 fingerprints of primitive biological activity.  In the coming decades many space and 
ground based facilities are planned that are designed to search for EPs
on all scales from massive, young “hot Jupiters”, through large rocky super-Earths 
down to the detection of exo-Earths within the “habitable” zone. 
Few of the planned facilities, however, will have the ability to 
characterize the planetary atmospheres which they discover through the application of MIR
and FIR spectroscopy.  The Japanese led SPICA space telescope will be realized within 10 years and 
has a suite of instruments that can be applied to the detection and characterization of 
EPs over the $\sim$5--210\,$\mu$m spectral range 
(see \textit{e.g.,} our \textit{White Paper}; Goicoechea et al. 2008).

\section{SPICA and the European Far-IR Instrument -- \textit{SAFARI}}  

\begin{figure}[ht]
  \centering 
  \includegraphics[height=0.4\hsize{},angle=0]{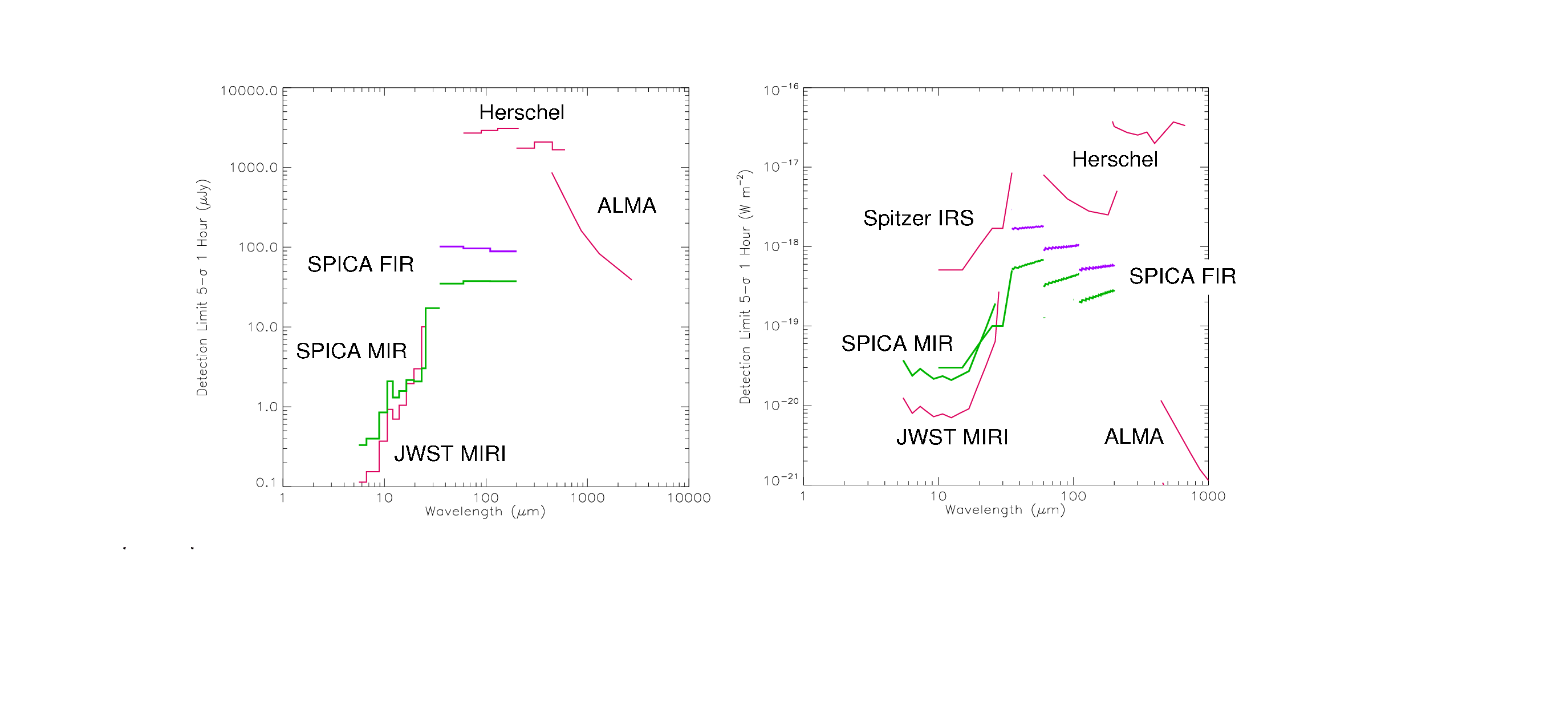} %
   \caption{\textit{Left}: Predicted photometric performance of SPICA,
   green (goal) and purple (requirement), compared to predecessor and complementary 
   facilities  given as point source sensitivities in $\mu$mJy for 
   5$\sigma$ in 1 hour over the bands shown indicatively as horizontal lines. 
   Note the 2 orders of magnitude increase in FIR photometric sensitivity compared 
   to \textit{Herschel} that will be achieved using goal sensitivity detectors on SPICA. 
   The figures here are raw sensitivity with no allowance for confusion.
   \textit{Right}: Same as previous for single unresolved line sensitivity 
   for a point source in W\,m$^{-2}$ (5$\sigma$--1\,hr). 
   For ALMA 100 km\,s$^{-1}$ resolution is assumed. The SPICA MIR sensitivities are
   scaled  by telescope area from the JWST and \textit{Spitzer}/IRS values respectively.}
  \label{fig:safari-sensitivity}
\end{figure}

SPICA will have the same size telescope as the ESA \textit{Herschel Space Observatory} 
(3.5\,m) but cooled to $<$5\,K \cite{swi08} thus removing its self emission and 
delivering an improvement in photometric sensitivity over \textit{Herschel} 
of two orders of magnitude 
in the FIR range (see Figure~\ref{fig:safari-sensitivity}). 
 The SPICA telescope will be monolithic, unlike the segmented JWST mirror, 
and will deliver diffraction limited performance at 5$\mu$m with a clean point 
spread function (PSF).

\textit{SAFARI} (SpicA FAR IR Instrument; Swinyard et al.) is a nationally 
funded FIR imaging spectrometer
proposed and developed primarily in Europe/Canada with possible contributions 
from Japan and US (NASA). 
\textit{SAFARI} will cover the FIR window that extends from $\sim$30\,$\mu$m 
(the upper cut-off of the MIR instruments) to $\sim$210\,$\mu$m (just longward of the 
[N\,{\sc ii}]206\,$\mu$m fine structure line) with a large field-of-view
 of $\sim$2$'$$\times$2$'$. 
Assuming diffraction limited performance, \textit{SAFARI} will provide angular resolutions 
from $\sim$2$''$ to $\sim$15$''$ (or $\sim$20 to $\sim$150\,AU at 10\,pc) at wavelengths not covered by 
JWST and, as shown in Figure~\ref{fig:safari-sensitivity}, at more than 2 orders of magnitude 
higher sensitivity than \textit{Herschel}/PACS.

The instrument concept that most closely matches the scientific requirements for 
large field-of-view, imaging and flexible spectral resolution is an imaging Fourier Transform
Spectrometer (FTS) in a \textit{Mach--Zehnder} configuration. Similar designs have been
already implemented in other ground-based and space telescopes
\cite{nay03, swi03}.
A spectral resolution of $R=\lambda/\Delta \lambda$=2000 at 100\,$\mu$m
(higher at shorter wavelengths) can be achieved
in the most simple and lightweight optical configurations.

\section{Exoplanet and Planet Formation Research in the Far-IR}  

\begin{figure}[h]
  \centering %
  \includegraphics[height=0.4\hsize{},angle=0]{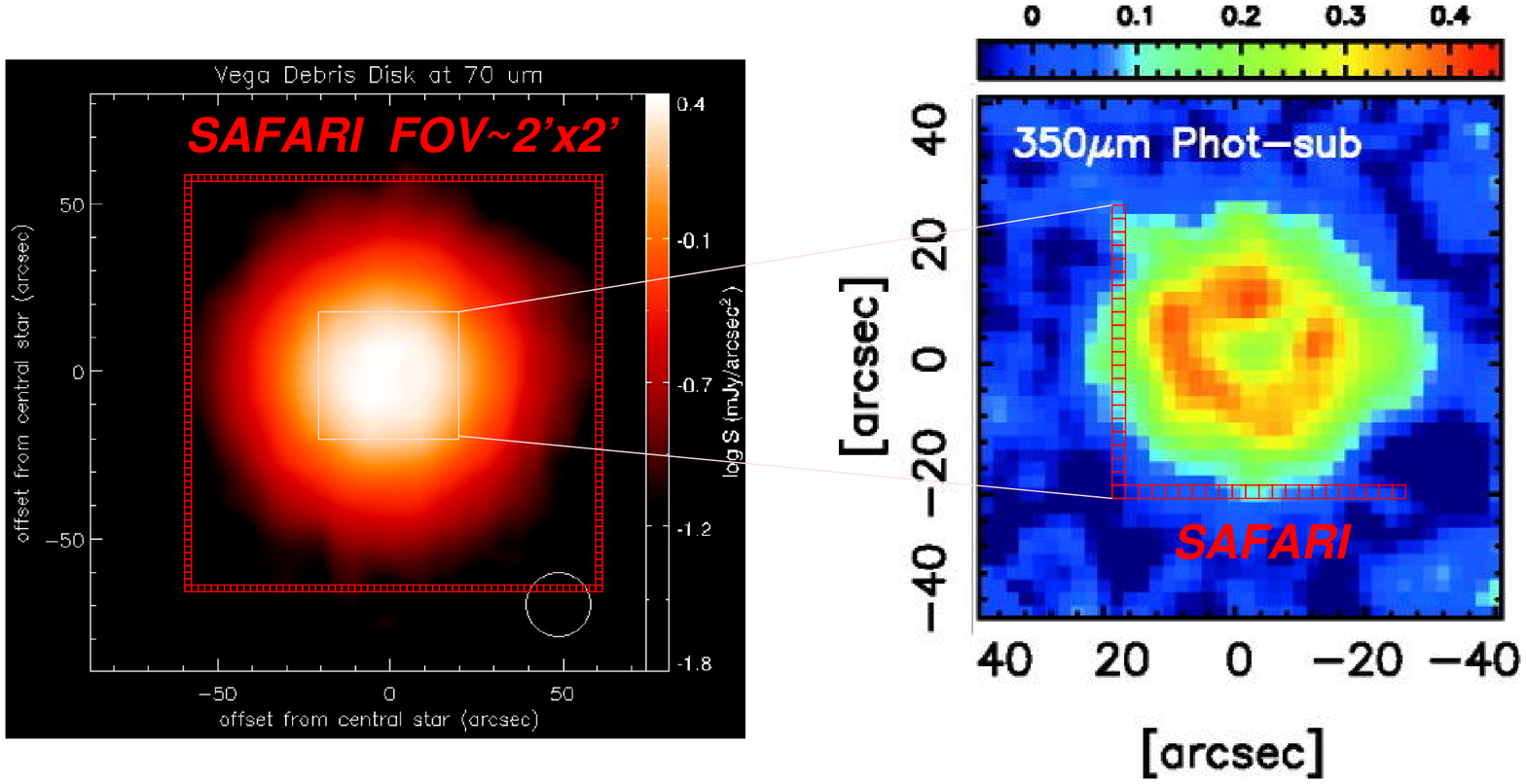} %
   \caption{\textit{Left}: Vega disk at 70\,$\mu$m observed with \textit{Spitzer}/MIPS. 
   The PSF is shown as a white circle \cite{su05}. \textit{SAFARI}'s 
   field-of-view and pixel 
   size ($\sim$1.8'') at 44\,$\mu$m (water ice feature) are shown with red squares.
   \textit{Right}:  CSO SHARC II 350\,$\mu$m image of Vega \cite{mar06}
   over plotted with the pixel scale of SPICA at 44\,$\mu$m (red squares). 
   SAFARI's spatial resolution is equivalent to ~23 AU at this distance and will allow
   to detect the presence of the expected ``snow''-free region of 42 AU in diameter.}
  \label{fig:fir-vega}
\end{figure}

 The huge increase in sensitivity compared to \textit{Herschel} can potentially open EP 
research to wavelengths completely blocked by the Earth’s atmosphere, but representing 
the emission peak of many cool bodies (gas-giant planets, asteroids and so on).  
\textit{SAFARI} will have its major strength in measuring excess radiation from dusty 
proto-planetary disks in hundreds of stars at almost all galactic distances.  
It will also perform medium spectral resolution observations over a spectral
 range rich in dust features, water vapor rotational lines (temperatures below $\sim$500 K), 
atomic oxygen fine structure lines at $\sim$63 and $\sim$145\,$\mu$m and  solid state 
water-ice features at $\sim$44 and $\sim$62\,$\mu$m.
Water ice is the major ingredient of the core of gas giant planets comets and smaller objects
 and its detection in proto planetary systems will
provide the observational evidence needed to constrain models of planetary formation.
In nearby systems \textit{SAFARI} will be able to image disks directly 
(see Figure~\ref{fig:fir-vega}) to examine their structure and trace the mineral and ice content
as a function of radius to compare them  with the spectra of comets, asteroids
and TNOs within our own Solar system.

\textit{SAFARI} will provide capabilities that complement SPICA's MIR coronagraph studies
(Enya et al. these proceedings) of extrasolar giant planets.  
First, it will be able to characterize the exozodiacal dust component of 
several thousand nearby stars. Determining the exo-zodiacal background levels from 
observations of a  large sample of stars will be key to prioritizing 
Earth-like candidates for searches with longer-term TPF-type missions 
(due to increased photon noise and potential confusion with zodiacal structures).   
With two orders of magnitude in sensitivity over \textit{Herschel}, SPICA would be able to 
survey stars to the level of dust mass that is limited by calibration accuracy 
i.e., 0.01 lunar mass for 90\,K grains around Sun-like stars \cite{wya08} out to 10 times 
greater distance (\textit{e.g.,} to 180\,pc as opposed to 18\,pc for Sun-like stars), 
resulting in $\sim$1000 times as many detections (\textit{e.g.,} $\sim$10$^5$ rather than 10$^2$ 
Sun-like stars can be surveyed with SPICA). 
SPICA will be particularly adept at studying both protoplanetary disks 
and the brief epoch at $\sim$10\,Myr at which the transition from protoplanetary to debris 
disk occurs.


\begin{figure}[h]
  \centering %
  \includegraphics[height=0.99\hsize{},angle=-90]{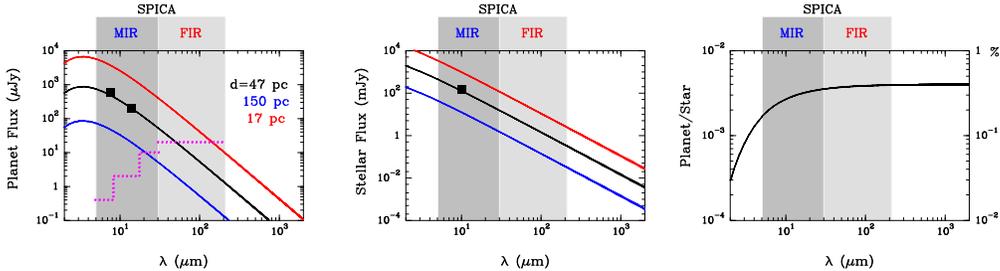} %
   \caption{\textit{Left}: Fit to HD 209458b hot Jupiter (T$_{eff}$$\sim$1000 K).
    MIR  fluxes were determined from \textit{Spitzer}/IRS observations \cite{swa08} 
    during its transit 
    around a G0 host star ($d$$\sim$47\,pc, T$_{eff}$$\sim$6000\,K) and estimations for 
   different distances. 
   The dashed lines show the estimated photometric sensitivity of SPICA with GOAL 
   detectors (5$\sigma$-1hr). \textit{Middle}: Expected flux from the host star. 
   \textit{Right}: Resulting planet-to-star contrast ratio. 
   The shaded regions represent the 2 main wavelength domains of SPICA instruments: 
  ``MIR'' for the coronagraph and MIR instruments and ``FIR'' for SAFARI.}
  \label{fig:fir-transit}
\end{figure}

Secondly, with very stable detectors and efficient, high cadence and high S/N observations, 
\textit{SAFARI} could also be used to perform transit photometry and, on the brightest candidates, 
spectroscopy for the first time in the FIR domain.  
FIR light curves will allow the characterization of EP radii as a function of wavelength, 
whilst occultation spectroscopy, during secondary eclipses, may reveal the same bright 
rotational water emission lines detected by ISO in the atmospheres of 
Jupiter, Saturn, Titan, Uranus and Neptune \cite{feu99}.  
Note that with \textit{SAFARI}'s sensitivities one could potentially extract the FIR 
 continuum of EPs like HD 209458b at distances out to $<$20\,pc 
(Figure~\ref{fig:fir-transit}).  
In addition, it may be possible to extract the thermal 
emission of nearby (massive-enough) cool planets (100-200 K) as a detectable contrast 
only occurs at wavelengths longer than $\sim$30\,$\mu$m.\\

\acknowledgements 

JRG is supported by a \textit{Ram\'on y Cajal} research contract from the 
Spanish MICINN and co-financed by the European Social Fund.

\end{document}